\documentclass{article}

\usepackage{authblk}
\usepackage{amssymb}
\usepackage{amsfonts}
\usepackage{amsmath}
\usepackage{algorithm}
\usepackage{algorithmic}
\usepackage{blindtext}
\usepackage{ifpdf}
\usepackage{graphicx}
\usepackage{cite}
\usepackage{color,soul}
\usepackage{balance}
\usepackage{caption}
\usepackage{subcaption}
\usepackage[colorlinks=true,linkcolor=blue,citecolor=blue]{hyperref}
\usepackage{apacite}
\usepackage[top=2.5cm, bottom=2.5cm, left=3cm, right=3cm]{geometry}
\usepackage{booktabs}
\usepackage[x11names]{xcolor}
\usepackage{soul}

\definecolor{blue}{rgb}{0, 0.188, 0.858}

\newif\ifboldnumber

\newenvironment{keywords}{\paragraph*{Keywords:}}

\title{Multilayer Perceptron Neural Network Model: A Novel Approach for LFP Contrast Sensitivity Tuning}
\date{}
\author[1,2]{\small Sahar Maleki}
\author[1]{\small Reza Lashgari \thanks{Corresponding author. Email: r\_lashgari@sbu.ac.ir}}
\author[3]{\small Mahdi Aliyari Shoorehdeli}
\author[2]{\small Mohammad Komareji}
\affil[1]{\footnotesize Institute of Medical Science and Technology, Shahid Beheshti University, Tehran, Iran}
\affil[2]{\footnotesize Department of Mechanical, Electrical and Computer Engineering, Science and Research Branch, Islamic Azad University, Tehran, Iran}
\affil[3]{\footnotesize Department of Electrical Engineering, K. N. Toosi University, Tehran, Iran}

\begin{document}
\baselineskip=1.6 \normalbaselineskip
\maketitle
\begin{abstract}
\baselineskip=1.6 \normalbaselineskip
\noindent
Local field potentials (LFPs) have been demonstrated to be an important measurement to study the activity of a local population of neurons. The response tunings of LFPs have been mostly reported as weaker and broader than spike tunings. Therefore, selecting optimized tuning methods is essential for appropriately evaluating the LFP responses and comparing them with neighboring spiking activity. In this paper, new models for tuning of the contrast response functions (CRFs) are proposed. To this end, luminance contrast-evoked LFP responses recorded in primate primary visual cortex (V1) are first analyzed. Then, supersaturating CRFs are distinguished from linear and saturating CRFs by using monotonicity index (MI). The supersaturated recording data are then identified through static identification methods including multilayer perceptron (MLP) neural network, radial basis function (RBF) neural network, fuzzy model, neuro-fuzzy model, and the local linear model tree (LOLIMOT) algorithm. Our results demonstrate that the MLP neural network, compared to traditional and modified hyperbolic Naka-Rushton functions, exhibits superior performance in tuning the local field potential responses to luminance contrast stimuli, resulting in successful tuning of a significantly higher number of neural recordings of all three types. These results suggest that the MLP neural network model can be used as a novel approach to measure a better fitted contrast sensitivity tuning curve of a population of neurons than other currently used models.
\end{abstract}

\vspace{0.2cm}
\section{Introduction}
The primate brain is the most complex and mysterious biological structure in the world, and scientists still have many unanswered questions about its function. In recent decades, scientists have made extensive efforts to understand the neuronal activity. Due to the activity of a large population of neurons, the LFP signal originating from neural populations is more difficult to interpret than single spiking neurons. Hence, an important part of these efforts has been focused on characterizing the signaling properties and functional significance of local field potentials within the cerebral cortex.

For improved comprehension of LFPs, it is crucial to apply an optimized neural tuning model to LFP signals and then compare these tuned LFPs with concurrently recorded single-unit spiking activity from adjacent neurons. Precisely, optimized functional models and analysis are needed to understand the interpretation of LFP signals and the neural circuits within the cortex \cite{einevoll2013modelling}. The optimal local field potential tuning is characterized by its high selectivity for specific features, such as orientation, direction, and luminance contrast \cite{lashgari2012response, land2013response, li2015mixing, jansen2015chromatic, jansen2019cortical}. Among the features, contrast sensitivity is considered one of the most important.

Prior studies have demonstrated a positive correlation between stimulus luminance contrast and LFP responses \cite{henrie2005lfp, peirce2007potential}. Nevertheless, LFP visual signals generally do not respond linearly to contrast. Instead, they often exhibit a saturated response to high-contrast stimuli. Furthermore, some LFP responses display supersaturation, decreasing at the highest contrast levels \cite{peirce2007potential, sani2013selective, hadjipapas2015parametric, lashgari2012response}. Cortical neuron and LFP responses to varying levels of contrast are commonly modeled with either the traditional Naka-Rushton function or its modified form \cite{peirce2007potential, lashgari2012response}. The modified Naka-Rushton model represents an improvement over the traditional version. Generally, this modified function effectively tunes neural luminance contrast responses for many neurons. However, even with this function, some neuron/LFP responses cannot be or are weakly tuned \cite{peirce2007potential}. Therefore, to overcome the shortcomings of the traditional and modified Naka-Rushton functions, a more accurate and higher-performing model for describing contrast-response functions is required.

Inspired by the aforementioned discussion, this paper proposes a novel approach for LFP contrast sensitivity tuning that overcomes the limitations of prior models. In this regard, single-unit spiking activity and local field potential signals, elicited by drifting sinusoidal gratings with eight luminance contrast levels $(0-76\%)$, were simultaneously recorded in the V1 of awake monkeys. The collected LFP signals are first analyzed. Next, the supersaturating CRFs are distinguished from linear and saturating CRFs by using monotonicity index. Afterwards, supersaturated recording data are identified utilizing a range of static identification methods, specifically MLP neural network, RBF neural network, fuzzy model, neuro-fuzzy model, and the LOLIMOT algorithm. The performance of these methods is then compared against the currently used models for all three data types. The main body of the paper prioritizes the MLP neural network approach, while supplementary details regarding the other methods are provided in the appendix. 

\section{Materials and Methods}\label{s2}
The following section details the methodologies and materials of the paper.

\subsection{Data Collection and Analysis}\label{sn2}
The data were collected as follows: monkeys (\textit{Macaca mulatta}, $n=3$ males) were implanted with a scleral eye coil, and the Eyelink was used to monitor eye movements. A head post was used for head fixation, and a chronic array of ultrathin electrodes protected by a recording chamber was implanted. In the third monkey, recordings were collected daily using a single FHC electrode in the primary visual cortex. Use of the macaque monkeys and the experimental protocol were approved by the Institutional Animal Care and Use Committee (IACUC) at the State University of New York, State College of Optometry and the Use Committee at the Institute for Research in Fundamental Sciences (IPM). The monkeys were housed and handled in strict accordance with the recommendations of the National Institutes of Health (NIH) protocols and health conditions were constantly monitored by the institutional veterinary doctor.

Visual stimuli were generated with a Matlab psychotoolbox and presented on a monitor (Sony Electronics; refresh rate, $160$ Hz; mean luminance, $61$ ${\rm{cd}}/{{\rm{m}}^2}$; resolution, $640 \times 480$ {\rm{pixels}}). The response properties quantified in this study were measured with drifting gratings presented for $2$ ${\rm{sec}}$ at $2$ Hz. After obtaining the feature properties of single isolated neurons; e.g., orientation and spatial frequency tunings, the stimuli were matched with the optimized response of neurons.  Contrast sensitivity was measured with eight different luminance contrasts ($0-76\%$). The monkeys were trained to touch a bar and fixate on a small cross ($0.1^\circ /{\rm{side}}$) presented at a distance of $57$ ${\rm{cm}}$. They maintained the eye position within $\pm 0.5 - {1^ \circ }\;$ for 3 ${\rm{sec}}$ and released the bar when the cross changed color to receive a drop of juice as reward. During the fixation, stimuli were presented at the receptive field location of the recorded single neuron to study the contrast tuning of the single neurons and LFPs.

Local field potential signals and single-unit spiking activity were simultaneously recorded in the V1 of awake monkeys. The electrodes were independently moved with individual microdrives, and the electrical signals from each electrode were collected and recorded \cite{lashgari2012response, fayyaz2019multifractal, jansen2019cortical}. A total of $66$ neurons in V1 of three monkeys were recorded by means of an array of brain-implanted electrodes, as well as through daily recordings with a single FHC electrode.

The electrical signals recorded from each neuron were amplified and digitized in eight contrast levels ($0-76\%$). For each neuron, the corresponding signals were subjected to the following steps: the signals were filtered either between $250$ Hz and $8$ kHz (spikes) or between $0.5$ Hz and $2.2$ kHz (LFPs) with two-pole low-cut and four-pole high-cut filters. Spikes and LFPs were sampled at $40$ and $5$ kHz, respectively, for two monkeys \cite{lashgari2012response} and were sampled at $30$ and $5$ kHz for the third monkey (Blackrock Microsystem). Then, for each contrast, the raw local field potential signals are low-pass filtered using a Butterworth filter (Second-order, $<100$ Hz). The local field potential signals are evaluated for stationarity, and a fast Fourier transform is used to convert LFP signals into the gamma band, defined as frequencies between 30 and 90 Hz. Next, for each contrast, the mean frequency power is calculated independently. Contrast response tuning function is normalized via dividing the mean frequency power of the stimulus-response by the mean frequency power of the baseline activity preceding the stimulus onset ($-200$ to $0$ ${\rm{ms}}$). Therefore, the units of the tuning functions are given as multiples of the signal-to-noise ratio (SNR or S/N) of the recordings, as described by
\begin{equation}\label{1}
{\rm{SN}}{{\rm{R}}_{{S_i}}} = \frac{{\mathop \sum \nolimits_{trial = {\rm{1}}}^n {{r_{{S_i}}} \mathord{\left/
				{\vphantom {{} {}}} \right.
				\kern-\nulldelimiterspace} {n}}}}{{\mathop \sum \nolimits_{trial = {\rm{1}}}^n {r_b} \mathord{\left/
			{\vphantom {{} {}}} \right.
		\kern-\nulldelimiterspace} {n}}}{},
\end{equation}
where ${r_{{S_i}}}\;$ is the mean frequency power of the response to the stimulus ${S_i}$, ${r_b}$  is the mean frequency power of the baseline activity, and $n$ is the number of stimulus trials \cite{lashgari2012response}. 

\subsection{Evaluation of Currently Used Models for Contrast Tunings}
In this subsection, the available grey-box models for contrast tunings including linear, Naka-Rushton, and modified Naka-Rushton models are described and evaluated. 

The linear model is described as follows
\begin{equation}\label{2}
R_{\rm C} = A.C + B,
\end{equation}
where $R_{\rm C}$ is the response to each contrast, $A$ and $B$ are unknown model parameters, and $C$ is the contrast stimulus \cite{albrecht1982striate}. 

The Naka–Rushton model is defined as
\begin{equation}\label{3}
R_{\rm C} = {R_{\rm m}}\frac{{{C^n}}}{{{C^n} + C_{50}^n}} + B,
\end{equation}
where $R_{\rm C}$ is the response to each contrast, and $B$ is the baseline activity. Also, ${R_{\rm m}}$ is the maximum response with baseline subtracted, $C$ is each stimulus contrast, ${C_{50}}$ is the contrast that generates the half-maximum response, and $n$ is the exponent that controls shape of function \cite{peirce2007potential, albrecht1982striate}.

The modified Naka–Rushton model, namely the Peirce function is described by
\begin{equation}\label{4}
R_{\rm C}= {R_{\rm m}}\frac{{{C^n}}}{{{C^{s.n}} + C_{50}^{s.n}}} + B,
\end{equation}
where the added parameter $s$ is the exponent that controls the function’s suppressive exponent \cite{peirce2007potential, lashgari2012response}.

Every recording is tuned by these models by using leave-one-out cross-validation (LOOCV) method in order to avoid overfitting the model. LOOCV is a particular case of $k$-fold cross-validation with $k=n$ ($n=8$ contrast). In $k$-fold cross-validation, the original sample is randomly partitioned into $k$ equal sized subsamples. Of the $k$ subsamples, a single subsample is retained as the test data for testing the model, and the remaining $k-1$ subsamples serves as training data. The cross-validation process is then repeated $k$ times. In each time, the model parameters obtained from the previous step are used as initial conditions for obtaining model parameters of the next step. 

The models are evaluated by using ${\rm{R}}^2$ \cite{celepcikay2009reg} and normalized mean square error (NMSE) \cite{poli1993use, alves2014evaluation} values described by Equations \eqref{5} and \eqref{6}. ${\rm{R}}^2$ is a statistic that will give some information about the goodness of fit of a model.
\vspace{0.3cm}
\begin{equation}\label{5}
{{\rm{R}}^2} = 1 - \frac{{{\rm{SSE}}}}{{{\rm{SST}}}} = 1 - \frac{{\mathop \sum \nolimits_{i = 1}^n {{\left( {{y_i} - {{\hat y}_i}} \right)}^2}}}{{\mathop \sum \nolimits_{i = 1}^n {{\left( {{y_i} - \bar y} \right)}^2}}},
\end{equation}
\vspace{0.3cm}
where ${\rm{SSE}}$ is sum of squares of errors, and ${\rm{SST}}$ is sum of squares of total. Also, ${{y_i}}$ denotes the output of the $i$-th contrast. Moreover, ${{{\hat y}_i}}$ represents the model output corresponding to the $i$-th contrast, ${\bar y}$ is mean outputs from eight contrasts, and $n$ is equal to $8$ that demonstrates eight contrasts.
\vspace{0.3cm}
\begin{equation}\label{6}
{\rm{NMSE}} = \frac{ {\mathop \sum \nolimits_{i = {\rm{1}}}^n {{(y_i-{\hat y_i})^2} \mathord{\left/
				{\vphantom {{} {}}} \right.
				\kern-\nulldelimiterspace} {n}}}}{{{{\bar y}}. {{\bar {\hat y}}} }},
\end{equation}
\vspace{0.3cm}
where $\bar {\hat {y}}$ is the average of the model outputs calculated across the eight luminance contrasts.

\begin{figure}[!h]
	\centering
	\includegraphics{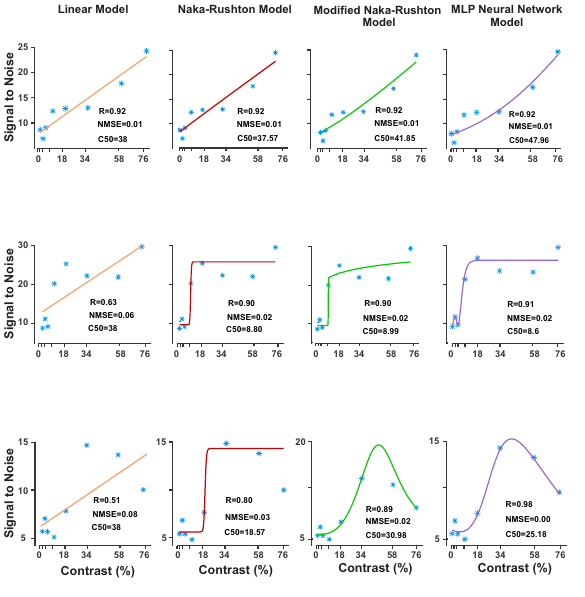}
	\caption{LFP contrast response tuning with linear and nonlinear models for three types of contrast sensitivity recordings including the linear (top), saturating (middle), and supersaturating (bottom) categories}
	\label{fig1}
\end{figure}

The mean evaluation results are shown in Table. \ref{tbl1}. Note that recordings with ${\rm{R}}^2 < 0.6$ are not included in the table.
\begin{table}[!t]
	\caption{Average goodness of fit (${\rm{R}}^2$) and average ${\rm{NMSE}}$ for tuned recording sites with ${\rm{R}}^2$$\ge 0.6$}\label{tbl1}
	\centering
	\begin{tabular}{|c|c|c|c|c} 
		\toprule
		\textbf{Model} & $\textbf{${\rm{R}}^2$}$  & $\textbf{{\rm{NMSE}}}$  & $N^{*}$ \\\hline
		\midrule
		Linear &$0.88$   &$0.02$    &$55 \quad (0.83\%)$  \\\hline
		Naka-Rushton  &$0.90$   &$0.01$    &$55 \quad (0.83\%)$   \\\hline
		Modified Naka-Rushton &$0.91$   &$0.01$    &$56 \quad (0.85\%)$  \\\hline
		MLP Neural Network&$\textbf{0.92}$   &$\textbf{0.01}$    &$\textbf{63  (0.95\%)}$ \\
		\bottomrule
	\end{tabular}

*$N$ represents the number of tuned recording sites where ${\rm{R}}^2$$\ge 0.6$.
\end{table}
For further consideration, a comparison between the performance of linear, Naka-Rushton, and modified Naka-Rushton models for three sample LFP recordings is illustrated in Figure \ref{fig1}. 
According to the evaluation results, the modified Naka-Rushton model is more accurate than linear and Naka-Rushton models, but it can also be improved. 

\subsection{Introducing a Model with Higher Performance than the Modified Naka-Rushton Function}\label{sn}
In this subsection, the objective is to present a model offering a higher performance compared to the modified Naka–Rushton. 

For this purpose, the supersaturating data are first separated from linear and saturating data at the gamma frequency band using the monotonicity index \eqref{7} by identifying data with ${\rm{MI}}$$<1$ as supersaturated.
\begin{equation}\label{7}
{\rm{MI}} = {\rm{1}} - \frac{{{R_{\rm m}} - {R_{{\rm{100}}}}}}{{{R_{\rm m }} - {R_{\rm{0}}}}},\;{\rm{\;\;\;\;\;\;}}
\end{equation}
where ${{R_{\rm m}}}$ is the maximum response of the LFP, ${{R_{{\rm{100}}}}}$ is the response at maximal contrast ($76$\%), and ${{R_{\rm{0}}}}$ is the response at the minimum contrast ($0$\%). This index takes a value of $1$ where the LFP's response rises monotonically and less than $1$ for LFPs demonstrating supersaturation. The LFPs whose response at the maximal contrast falls fully back to baseline rate takes an ${\rm{MI}}$ of $0$ \cite{peirce2007potential}.

Then, considering that the studied data were sampled statically, the effort is made to identify the system (supersaturated data) using static methods. The system is identified by a feedforward MLP neural network comprising a hidden and an output layer. The Sigmoid transfer function is used at the hidden layer, whereas the linear transfer function is applied to the output layer. Therefore, a structure in the form of Equation \eqref{8} is assumed to express the relationship between the system input and output \cite{nelles2001nonlinear, hagan2002neural}.
\begin{equation}\label{8}
y = \mathop \sum \limits_{i = 1}^n \frac{{{\alpha _i}}}{{1 + {e^{ - \left( {{w_i}\phi  + {b_i}} \right)}}}} + {\alpha _0} + e,
\end{equation}
where $\phi $ is the input, ${{w_i}}$, ${{b_i}}$, ${{\alpha _i}}$, and $n$ are the model parameters, $e$ represents the error value, and $y$ is the system output.
Note that the system input data are normalized between $0$ and $1$ by Equation \eqref{12}.
\begin{equation}\label{12}
	\begin{aligned}
		{\phi _N} = \frac{{\phi  - {\phi _{min}}}}{{{\phi _{max}} - {\phi _{min}}}},
	\end{aligned}
\end{equation}
where ${{\phi _{min}}}$ shows the minimum input, ${{\phi _{max}}}$ is the maximum input, $\phi$ represents the input data, and ${{\phi _{N}}}$ is the normalized input.

Parameters ${\alpha _i};i = 0,1, \ldots ,n$, ${w_i};i = 1, \ldots ,n$, and ${b_i};i = 1, \ldots ,n$ are estimated to minimize the cost function presented in Equation \eqref{9} by the nonlinear Levenberg–Marquardt optimization using Equation \eqref{10} and the error backpropagation learning algorithm.
\begin{equation}\label{9}
\begin{aligned}
J &= \frac{1}{2}\mathop \sum \limits_{k = 1}^N {\left( {e\left( k \right)} \right)^2}= \frac{1}{2}\mathop \sum \limits_{k = 1}^N {\left( {y\left( k \right) - \left( {\mathop \sum \limits_{i = 1}^n \frac{{{\alpha _i}}}{{1 + {e^{ - \left( {{w_i}\phi \left( k \right) + {b_i}} \right)}}}} + {\alpha _0}} \right)} \right)^2}
\end{aligned}
\end{equation}
\begin{equation}\label{10}
\begin{aligned}
{\hat \theta _{i + 1}} = {\hat \theta _i} - {\left( {\mathop \sum \limits_{k = 1}^N \frac{{\partial \hat y\left( {k.{{\hat \theta }_i}} \right)}}{{\partial {{\hat \theta }_i}}}{{\left( {\frac{{\partial \hat y\left( {k.{{\hat \theta }_i}} \right)}}{{\partial {{\hat \theta }_i}}}} \right)}^T} + {\mu _i}I\;} \right)^{ - 1}}\nabla J\left( {{{\hat \theta }_i}} \right),
\end{aligned}
\end{equation}
\vspace{0.1cm}
in which
\vspace{0.1cm}
\begin{equation*}\label{11}
	\begin{aligned}
		\nabla J\left( {{{\hat \theta }_i}} \right) =  - \mathop \sum \limits_{k = 1}^N \left( {y\left( k \right) - \hat y\left( {k.\theta } \right)} \right)\frac{{\partial \hat y\left( {k.\theta } \right)}}{{\partial \theta }}\left| {\theta  = {{\hat \theta }_i}} \right.,
	\end{aligned}
\end{equation*}
where ${{{\hat \theta }_i}}$ represents the estimated network parameters ($\theta$) after the $i$-th iteration of the optimization method, ${\mu _i} > {\rm{0}}$ is the step size, $\nabla J\left( {{{\hat \theta }_i}} \right)$ is the gradient vector of the cost function $J$ at the point ${{{\hat \theta }_i}}$, and $N$ is the size of the training dataset.

The optimal number of neurons in the hidden layer ($n$) and the optimal number of epochs are determined based on validation data to prevent the model overfitting on the training data. To this end, first, for a fixed number of epochs, the number of neurons is changed, and the resulting mean squared error (MSE) is plotted in multiple iterations based on validation and training data. In each plot, the number of neurons corresponding to a simultaneous increase in the mean squared error of the validation data and a decrease in that of the training data is accepted as the optimal value. The average and variance of the results for the optimal number of neurons are obtained based on multiple iterations of the program. Furthermore, to determine the optimal number of epochs, the same process is repeated using the optimal number of neurons already obtained.

\section{Results}\label{s3}
\begin{figure*}[!t]
	\centering
	\includegraphics[width=\textwidth]{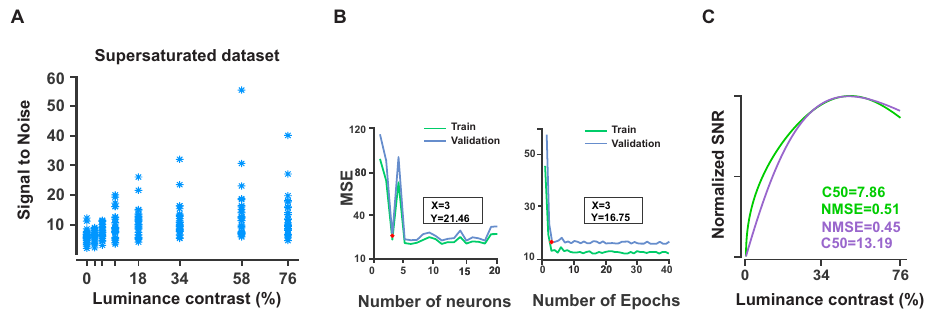}
	\caption{Contrast tuning model proposed for tuning the supersaturated LFP dataset (purple) in comparision with the modified Naka-Rushton function (green). \textbf{A.} The LFP supersaturated dataset. \textbf{B.} One run for specifying the optimal number of neurons (Left) and epochs (Right) for the MLP neural network model. \textbf{C.} The MLP-neural-network-based model of Equation \eqref{13} with parameters estimated by the nonlinear Levenberg–Marquardt optimization and the backpropagation learning method.}
	\label{fig2}
\end{figure*}

As mentioned in Subsection \ref{sn2}, in this study, single-unit spiking activity and local field potential signals, elicited by drifting sinusoidal gratings with eight luminance contrast levels $(0-76\%)$, were simultaneously recorded in the V1 of awake monkeys. After data processing and stationary evaluation, the fast Fourier transform is used to study the gamma LFP frequency band ($30-90$ Hz). Further analysis is done on the SNR of the recordings. According to Subsection \ref{sn}, the supersaturated recordings are distinguished from linear and saturated recordings using the monotonicity index defined in Equation \eqref{7}; MI smaller than 1 (${\rm{MI}}$$<1$) is regarded as supersaturated and MI equals 1 (${\rm{MI}}$$=1$) is known as linear and saturating LFP recordings. After proceeding with these steps, $224$ ($28 \times 8$) data points as the supersaturated depicted in Figure \ref{fig2}{\color{blue}A} and $304$ ($38 \times 8$) as the linear and saturated data points are identified.
Then, the $224$ supersaturated data points are divided into $156$ training, $34$ validation, and $34$ test data points. The studied system is identified by multilayer perceptron neural network. In this model, the parameters are estimated by the nonlinear Levenberg–Marquardt optimization and the error backpropagation learning method. For this analysis, the system input data are normalized between $0$ and $1$ by Equation \eqref{12}, and the initial values of weights and deviations are randomly selected between $-0.6$ and $0.6$. Also, the initial training step is considered to be $0.01$. Figure \ref{fig2}{\color{blue}B} shows the results of MLP neural network from one execution of the algorithm to determine the optimal number of neurons and the optimal number of epochs, respectively. Based on the optimal number of neurons and epochs obtained in $50$ iterations of running the program, the average and variance of the results for the optimal number of neurons are estimated as $2.64{\rm{ }} \pm 1.74$, whereas those of the optimal number of epochs are estimated as $3.24{\rm{ }} \pm 2.26$. Considering three neurons in the hidden layer, the model is obtained as 
\begin{equation}\label{13}
	\begin{aligned}
		\hat y = \mathop \sum \limits_{i = 1}^3 \frac{{{\alpha _i}}}{{1 + {e^{ - \left( {{w_i}\phi \left( k \right) + {b_i}} \right)}}}} + {\alpha _0},
	\end{aligned}
\end{equation}
the parameters of which are estimated by the nonlinear Levenberg–Marquardt optimization and the error backpropagation learning method by three epochs. This results in the NMSE value of $0.45$ for the test data. The modified Naka–Rushton model parameters are also estimated with $156$ training data points and then evaluated using $34$ test data points. For a better comparison of the presented model with the modified Naka–Rushton, the tunings are plotted for each one and are normalized for comparison in Figure \ref{fig2}{\color{blue}C}.

The other methods are also proposed and compared with modified Naka-Rushton function in appendix.

\section{Discussion}\label{s4}
\begin{figure*}[!tp]
	\centering
	\includegraphics[height=4.7cm]{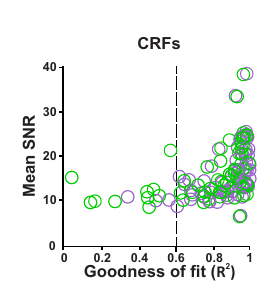}
	\caption{Goodness of fit for different LFP CRFs: performance comparison of the modified Naka-Rushton function (green) and MLP neural network model (purple)}
	\label{fig3}
\end{figure*}

The results demonetrate that the MLP neural network can be the best model for tuning LFP contrast response functions. According to the NMSE and ${C_{50}}$ values presented in Figure \ref{fig2}{\color{blue}C}, the MLP neural network shows a reliable contrast response and is comparable with the modified Naka-Rushton model. For better consideration, the mean goodness of fit (${\rm{R}}^2$) and average ${\rm{NMSE}}$ for the recording sites tuned with the MLP neural network have also been added to Table \ref{tbl1}. Also, for the three sample LFP recordings of Figure \ref{fig1}, contrast response functions have also been tuned with the MLP model and added to this figure. According to Table \ref{tbl1} and Figure \ref{fig1}, the MLP model has a better performance for all three types of LFP recordings than the three pre-existing models. To further investigate the performance of the model, Figure \ref{fig3} shows the distribution of goodness of fit values for all recordings, comparing the performance of the MLP neural network model to that of the modified Naka-Rushton model.
Based upon the results obtained and the discussion presented, the multilayer perceptron neural network function performs a better LFP contrast tuning for both linear and nonlinear (saturating and supersaturating) CRFs. This suggests that the MLP can be used as a novel approach to measure the LFP contrast sensitivity tunings.

\section{Conclusion}\label{s5}
In this paper, new kernel functions are introduced for improving careful measurements of LFP contrast-response tunings, and they are compared with the existing contrast-response models commonly found in the literature. To accomplish this, the collected LFP recordings are analyzed, and the supersaturating CRFs are distinguished from linear and saturating CRFs. Then, different models are presented by static methods for supersaturated data. Finally, all identified models are compared with the existing contrast-response models. It is shown that all identified models exhibit a higher performance compared to the modified Naka–Rushton function for supersaturated data. However, based on the criterion of avoiding overfitting, the MLP neural network model is suggested as the best model among the identified models for all three types of contrast sensitivity signals. Moreover, when evaluating models based on their performance, the MLP neural network demonstrates the most effective representation of all three kinds of contrast sensitivity signals, exhibiting the minimum NMSE and the highest goodness of fit in comparison to linear, Naka-Rushton, and modified Naka-Rushton models. Therefore, our results show that the multilayer perceptron neural network results in a better performance than the contrast tuning functions most commonly used.

\appendix 
\section{Appendix} 
In this section, other identified models are proposed and evaluated. These methods can be followed in \cite{norton2009introduction, chen1989orthogonal, jang1993anfis, zadeh1996fuzzy, takagi1985fuzzy, babuvska2003neuro, jang1995neuro}.

\vspace{0.2cm}
\subsection{RBF Neural Network for LFP Tuning}
\begin{figure*}
	\centering
	\includegraphics{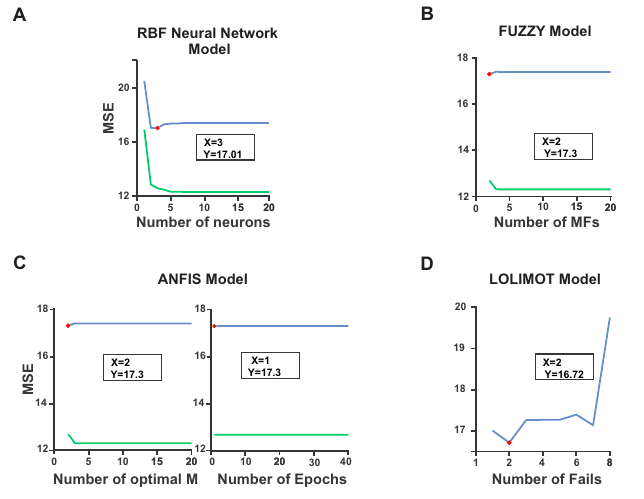}
	\caption{The determination of the optimal value of parameters for the new models.  \textbf{A.} Optimizing the number of neurons in RBF neural network model. \textbf{B.} Optimizing the number of rules in fuzzy model. \textbf{C.} Optimization of M (Left) and the number of epochs (Right) in ANFIS model. \textbf{E.} Optimization of $nl$ in LOLIMOT algorithm.}
	\label{fig5}
\end{figure*}

\begin{figure*}
	\centering
	\includegraphics{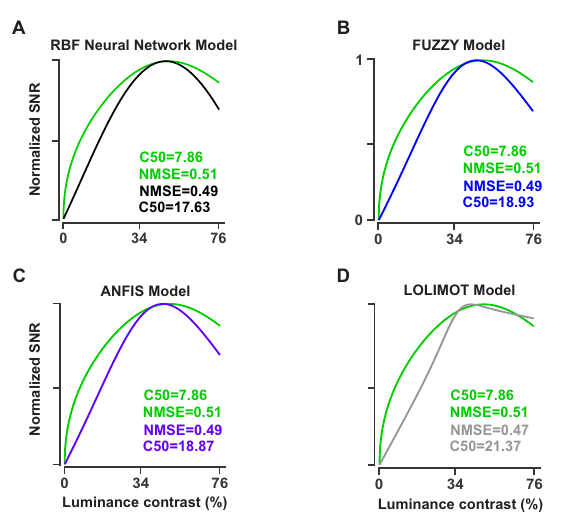}
	\caption{The proposed contrast tuning models vs. the modified Naka-Rushton function (green) for fitting supersaturated LFP responses. \textbf{A.} The RBF-neural-network-based model of Equation \eqref{14} with its estimated parameters. \textbf{B.} The fuzzy model of Equation \eqref{15} with its estimated parameters. \textbf{C.} The ANFIS model of Equation \eqref{16} with its estimated parameters. \textbf{D.} The model of Equation \eqref{17} identified using the LOLIMOT algorithm with its estimated parameters.}
	\label{fig6}
\end{figure*}

In this function, the model parameters are estimated using the forward selection orthogonal least squares method. The ${\sigma _i}$ is set to 38, selected as the best number for ${\sigma _i}$ for all values of $i$. Figure \ref{fig5}{\color{blue}A} illustrates the optimization of the number of neurons using the RBF neural network.
According to the figure, the optimal number of neurons is obtained as 3. Considering three neurons in the hidden layer, the parameters of the RBF neural network defined in Equation \eqref{14} are estimated using the forward selection orthogonal least squares method.
 
\begin{equation}\label{14}
\begin{aligned}
\hat y = \mathop \sum \limits_{i = 1}^3 {\alpha _i}{e^{\left( { - \frac{{{{\left( {\phi  - {c_i}} \right)}^2}}}{{2{\sigma _i}^2}}} \right)}} + {\alpha _0}
\end{aligned}
\end{equation}
The model is then evaluated using the test data, for which the NMSE is obtained a value of $0.49$, as shown in Figure \ref{fig6}{\color{blue}A}.

\vspace{0.3cm}
\subsection{Fuzzy Model for LFP Tuning}
In the fuzzy model, the parameters are estimated by network partitioning and linear least squares method. Figure \ref{fig5}{\color{blue}B} illustrates the optimization of the number of rules using fuzzy model. According to Figure \ref{fig5}{\color{blue}B}, the optimal number of rules is found to be 2. After estimating the parameters of the fuzzy model of Equation \eqref{15}, the model is evaluated using the test data, and a value of 0.49 is obtained for the NMSE, as depicted in Figure \ref{fig6}{\color{blue}B}.
\begin{equation}\label{15}
\begin{aligned}
\hat y = \frac{{\mathop \sum \nolimits_{i = 1}^2 \left( {{a_i}.\phi \left( k \right) + {b_i}} \right).{e^{\left( { - \frac{{{{\left( {\phi  - {c_i}} \right)}^2}}}{{2{\sigma _i}^2}}} \right)}}}}{{\mathop \sum \nolimits_{i = 1}^2 {e^{\left( { - \frac{{{{\left( {\phi  - {c_i}} \right)}^2}}}{{2{\sigma _i}^2}}} \right)}}}}
\end{aligned}
\end{equation}

\vspace{0.3cm}
\subsection{Neuro-Fuzzy Model for LFP Tuning}
The model parameters are estimated by the gradient descent optimization method applied to the loss function, the backpropagation learning rule, and the linear least squares method. The initial values of ${c_i}$ and ${\sigma _i}$ are determined by network partitioning. A training step of $0.01$ is also selected. Figure \ref{fig5}{\color{blue}C} illustrates the optimization of the number of rules $(M)$ and epochs, respectively. Considering an $M$ value of 2 and one epoch and estimating the neuro-fuzzy model parameters of Equation \eqref{16}, the model is evaluated using the test data, resulting in an NMSE of 0.49 as shown in Figure \ref{fig6}{\color{blue}C}.
\begin{equation}\label{16}
\begin{aligned}
\hat y = \frac{{\mathop \sum \nolimits_{i = 1}^2 \left( {{a_i}.\phi \left( k \right) + {b_i}} \right).{e^{\left( { - \frac{{{{\left( {\phi  - {c_i}} \right)}^2}}}{{2{\sigma _i}^2}}} \right)}}}}{{\mathop \sum \nolimits_{i = 1}^2 {e^{\left( { - \frac{{{{\left( {\phi  - {c_i}} \right)}^2}}}{{2{\sigma _i}^2}}} \right)}}}}
\end{aligned}
\end{equation}

\vspace{+0.3cm}
\subsection{LOLIMOT Algorithm for LFP Tuning}
The system is also identified by the LOLIMOT algorithm. Figure \ref{fig5}{\color{blue}D} illustrates the optimization of the number of local models ($nl$) using the LOLIMOT algorithm. Considering an $nl$ value of 2 and estimating the parameters of Equation \eqref{17}, the model is evaluated using the test data, resulting in an NMSE of 0.47 based on Figure \ref{fig6}{\color{blue}D}.
\begin{equation}\label{17}
\begin{aligned}
\hat y = \frac{{\mathop \sum \nolimits_{i = 1}^2 \left( {{a_i}.\phi \left( k \right) + {b_i}} \right).{e^{\left( { - \frac{{{{\left( {\phi  - {c_i}} \right)}^2}}}{{2{\sigma _i}^2}}} \right)}}}}{{\mathop \sum \nolimits_{i = 1}^2 {e^{\left( { - \frac{{{{\left( {\phi  - {c_i}} \right)}^2}}}{{2{\sigma _i}^2}}} \right)}}}}
\end{aligned}
\end{equation}

\subsection{An In-depth Evaluation of the Identified Models}
\begin{figure*}
	\centering
	\includegraphics{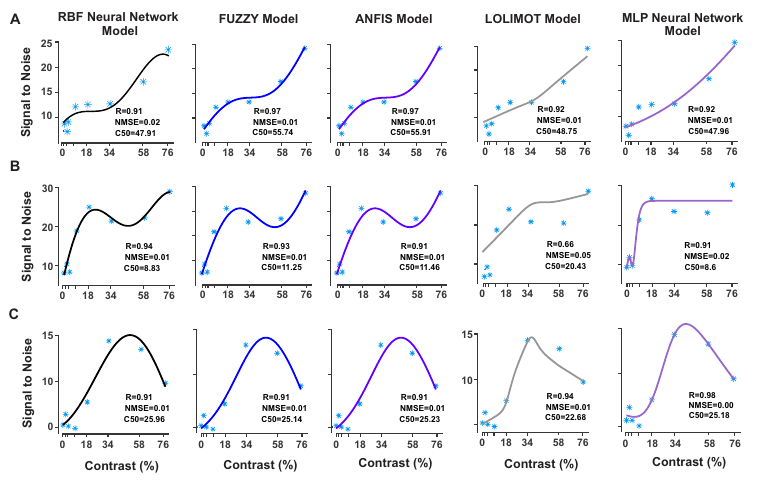}
	\caption{Comparison of different functional models for tuning LFP contrast response functions (an extension of Figure \ref{fig1})}
	\label{fig7}
\end{figure*}

For a more detailed examination of the identified models, contrast response functions of Figure \ref{fig1} are tuned with all identified models and depicted in Figure \ref{fig7}. As shown, all identified models exhibit a higher performance compared to the modified Naka–Rushton function for supersaturated data. However, considering the avoidance of overfitting as the primary criterion for model selection, the MLP neural network is the best tuning function for all three types of LFP contrast sensitivity signals.

\bibliographystyle{apacite}
\bibliography{Refs}

\end{document}